\documentclass[10pt,letterpaper]{article}
\usepackage[top=0.85in,left=0.75in,footskip=0.75in]{geometry}

\usepackage{changepage}
\usepackage[utf8]{inputenc}
\usepackage{textcomp,marvosym}
\usepackage{fixltx2e}
\usepackage{amsmath,amssymb}
\usepackage{cite}
\usepackage{nameref,hyperref}
\usepackage[right]{lineno}
\usepackage{microtype}
\DisableLigatures[f]{encoding = *, family = * }
\usepackage{rotating}

\usepackage[aboveskip=1pt,labelfont=bf,labelsep=period,justification=raggedright,singlelinecheck=off]{caption}

\makeatletter
\renewcommand{\@biblabel}[1]{\quad#1.}
\makeatother

\date{}

\usepackage{lastpage,fancyhdr,graphicx}
\usepackage{epstopdf}

\begin{document}
\vspace*{0.35in}

\begin{center}
{\Large
\textbf\newline{Pattern Formation in Populations with Density-Dependent Movement and Two Interaction Scales}
}
\newline
\\
Ricardo Mart\'inez-Garc\'ia\textsuperscript{1,2,*},
Clara Murgui\textsuperscript{1},
Emilio Hern\'andez-Garc\'ia\textsuperscript{1},
Crist\'obal L\'opez\textsuperscript{1}

\bigskip
\bf{1} IFISC (Instituto de F\'isica Interdisciplinar y Sistemas Complejos), CSIC-UIB, Campus UIB, Palma de Mallorca, Spain.
\\
\bf{2} Department of Ecology and Evolutionary Biology, Princeton University, Princeton NJ, 08544-1003, USA.
\\
\bigskip

* ricardom@princeton.edu

\end{center}
\section*{Abstract}
We study the spatial patterns formed by a system of interacting particles where
the mobility of any individual is determined by the population crowding at two different spatial scales. In this way we model the behavior
of some biological organisms (like mussels) that tend to cluster at short ranges
as a defensive strategy, and strongly disperse if there is a high population
pressure at large ranges for optimizing foraging.
We perform stochastic simulations of a particle-level model of the 
system, and derive and analyze a continuous density description (a 
nonlinear diffusion equation). In both cases we show that
this interplay of scale-dependent-behaviors gives rise to a rich
formation of spatial patterns ranging from labyrinths to periodic cluster arrangements. In most cases 
these clusters have the very peculiar appearance of ring-like structures, i.e., organisms arranging in the perimeter of
the clusters, that we discuss in detail.


\section*{Introduction}
Individual based models are of great relevance in many
disciplines, ranging from condensed matter \cite{Krapivsky2010}
to biology \cite{Murray2002a,Okubo2001}, economics and social
dynamics \cite{opinion}. They allow to simulate some simple
dynamical rules and study its consequences at a global
population scale.
In an ecological
context individual based models have gained in importance
\cite{grimm2013individual,judson1994rise,deangelis2005individual}, and
are commonly used to investigate collective behavior and the emergence of patterns,
which are central issues in theoretical ecology \cite{levin1992problem}.

In this paper, we propose a model to study the formation of
spatial patterns in a population of organisms in which
interactions affect their mobility. We assume that, during the
time scales of interest here, no other dynamical processes such
as birth, reproduction and death occur. The movement of any
individual depends on the distribution of its conspecifics at
two length scales. We thus focus on the problem of group
formation and spatial aggregation
\cite{Young2001,medvinsky2002spatiotemporal,Hernandez-Garcia2004,Heinsalu2010}, although
this approach may be used in the more general context of
collective movement \cite{vicsek2012collective} (including birds
flocks \cite{COUZIN2002,Bialek2012}, fish swarms
\cite{Ward2008,Lopez2012}, and mammals herds
\cite{Martinez-Garcia2013b}), and also to address the effect of
spatial degrees of freedom in evolutionary problems
\cite{deangelis2005individual}.

Spatial aggregation is a widespread phenomenon in living
systems, resulting from the combination of individual movement
with interspecific and intraspecific interactions
\cite{krause2002living,Okubo2001}. Therefore, a  mathematical
description of group formation should include all these
mechanisms, and several ways of integrating collective
interactions with individual movement have been proposed
\cite{Flierl1999,Murray2002a,viswanathan2011physics,
vicsek2012collective,marchetti2013hydrodynamics,mendez2013stochastic}.
A very important type of models considers that interactions
influence only the movement of the individuals disregarding any
other intra- and inter-specific interactions. They are relevant
to study animal or organism dispersal wherever there is an
increase of the diffusivity with the local density because of
population pressure \cite{Murray2002a,Okubo2001}. Extensions of
these works also account for the effect of conspecifics located
at separated positions
\cite{Burger2012,Murray2002a,Flierl1999}, including nonlocal
spatial interactions. This family of models results in nonlocal
nonlinear diffusion equations \cite{Lopez2006,Mendez2012} for
the population density. From a biological point of view, they
usually account for a single class of interactions, and
diffusivity depends on the population density over one
neighborhood of the focal individual. However, in a more
general framework, many different interactions of diverse
nature are relevant within a population, so these single-scale
approaches might not describe the complete set of processes
taking place. For instance, high long-range densities (i.e.
densities of others within a long distance around a focal one)
may encourage animal mobility due to intraspecific competition
for resources, while on a shorter spatial scale individuals may
arrange in cooperative aggregations so that the predation risk
decreases. Also in the decision-making process that underlies
collective movement, animals choose how to move depending on
their neighbors at different distances, so they guarantee the
cohesion of the group
\cite{COUZIN2002,Raghib2010,Schellinck2011}. In a rather
different context,  the formation of patterns of vegetation has
been also traditionally thought to be a consequence of the
interplay between plant interactions at two different scales:
short-range facilitation and long-range competition
\cite{Lefever1997,Borgogno2009,MartinezGarcia2013156,rietkerk2002self,bonachela2015termite},
although this has been a contentious claim
\cite{ptrsamartinez2014}).

Mussel beds are one of the paradigmatic examples of spatial
aggregation in nature. 
Experimental works have shown that the origin of the aggregates lies in the interaction among
individuals \cite{VandeKoppel2008}, although modified by the interplay between the whole 
population and the environment \cite{DeJager2011}. Many theoretical attempts have proposed
mathematical models to unveil the mechanisms that, acting at different spatial scales, 
stabilize the patterns. Two families of models have arisen, 
both of them containing competition for resources
on a large scale and facilitation (aggregation promotion to diminish wave stress and predation risk) on a short scale: 
a) studies considering the dynamics of two populations (the algae and the mussels) 
\cite{VandeKoppel2005,liu2012,wang2009nonlinear};
and b), an study that 
focuses only on the dynamics of the population of mussels (unique species model), 
including the interaction with the environment (i.e, algae)
in nonlocal spatial terms \cite{Liu2013}.

Within this framework, but mainly motivated by \cite{Liu2013},
we present a model of interacting particles where the mobility
of the individuals, i.e. its diffusivity, depends on two
spatial scales. Movement is encouraged when the density is high
in a long-range (competition), and inhibited if it is so in a
short-range (i.e., cooperative aggregations are favored at
shorter scales). The principal novelty of our work with respect to 
\cite{Liu2013} is the Brownian nature of the motion of the
particles
in the discrete description of the
system and its generality, that allows the exploration of different relationships between
the diffusivity and the density of individuals. We will perform a numerical study of this 
stochastic picture and compare the results with the equivalent deterministic population level
approach.

In the following sections pattern formation will be studied
combining numerical and analytical techniques both in the
discrete-particle dynamics and its continuous-field density
equation.

\section*{Materials and Methods}
\subsection*{Individual-based dynamics}
Let us consider a population of $N$ individuals undergoing
Brownian movement with a diffusion coefficient that depends on
the densities of conspecifics at two separated length-scales: a
mean density $\tilde{\rho}_s$ at short range, $R_s$, and a mean
density $\tilde{\rho}_l$ over a long one,  $R_l$
($R_{s}<R_{l}$). We will denote the position of each particle
by ${\bf r}_i=(x_i, y_i)$ at any time $t$ in a two-dimensional
square system of lateral extent $L$ with periodic boundary
conditions.

\begin{figure}
\centering
\includegraphics[width=0.35\textwidth]{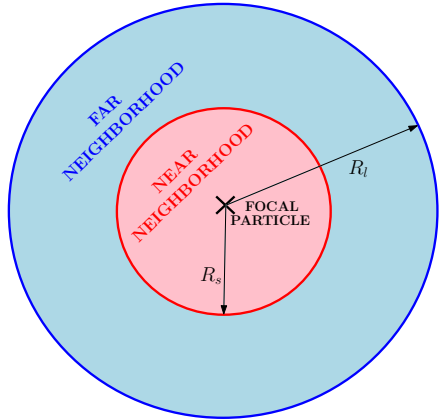}
\caption{{\bf Interaction neighborhoods.} Short- and long-range interaction neighborhoods for a given individual. 
The regions are defined by their radius $R_s$ and $R_l$ respectively.}
\label{fig1}
\end{figure}

The dynamics of each particle $i=1,\ldots N$ is then given by
\begin{equation}\label{langevin}
 {\bf \dot{r}}_{i}=\sqrt{2D \left({\bf r}_{i}, \tilde{\rho}_{s},{\tilde{\rho}_{l}}\right)} \boldsymbol{\eta}_{i}(t),
\end{equation}
where the diffusivity $D$ is, in general, a positive continuous
function of $\tilde{\rho}_{l}$ and $\tilde{\rho}_{s}$.
$\boldsymbol\eta_{i}(t)$ is a white Gaussian vector noise with
zero mean and with time-correlation matrix given by
$\langle\boldsymbol\eta_{i}(t) \boldsymbol
\eta_{j}(t')\rangle=\mathbf{1}\delta_{ij}\delta(t-t')$.
$\mathbf{1}$ is the identity matrix. Eq.~(\ref{langevin})
should be interpreted within the It\^{o} calculus, since the
stop/movement behavior is assumed to occur at the beginning of
each time step \cite{Lopez2006}. The mean densities are defined
as:
\begin{equation}
 \tilde{\rho}_{\mu}({\bf r})=\frac{N_{\mu}}{\pi R^{2}_{\mu}},
\end{equation}
with $\mu\equiv s,l$. $N_{s}$ and $N_{l}$ are the number of
individuals found in a near and far neighborhood of the
particle at $\bf r$, respectively (see Fig.~\ref{fig1}). Note
that, since the number of individuals does not change, the
global density $\rho_0=N/L^2$ remains constant in time.

The main focus of this approach is on species that switch
between the tendency to aggregation and to dispersion as the
number of surrounding individuals increases at different length
scales. In particular, we address the case of competitive long
range interactions and facilitation at a shorter scale. This is
the observed framework in mussel beds, where patterns
 appear due to the interaction between two opposing mechanisms:
competition for resources at a large scale and defense against
 predators and waves stress at shorter distance \cite{VandeKoppel2005,Liu2013,DeJager2011}

To model this behavior we consider that the diffusivity, $D$,
is enhanced with increasing the long-range density, and reduced
with increasing short-range density. This can be written as
$D=D_0 g(a-b \tilde{\rho}_{s} + c \tilde{\rho}_{l})$ if $g$ is
an arbitrary function with positive derivative,
$\partial_xg(x)>0$. $D_0$ is a constant diffusivity and $a$,
$b$, and $c$ are positive parameters. Note that with the
expression $a-b \tilde{\rho}_{s} + c \tilde{\rho}_{l}$ we
indicate, as mentioned before, that the {\it tendency} of a
particle to move decreases with the short-range mean density
($-b \tilde{\rho}_{s}$) and increases with the long-range one
($c \tilde{\rho}_{l}$). The function $g$ takes its maximum
(minimum) value in the limit
$\tilde{\rho}_{l}\gg\tilde{\rho}_{s}$
($\tilde{\rho}_{s}\gg\tilde{\rho}_{l}$). For simplicity we
restrict to the case  $0\leq g \leq 1$ so the diffusivity of
the particles varies between $0$ and $D_{0}$. $D_0$ is the
diffusion coefficient of the population when movement is
extremely promoted ($\tilde{\rho}_{l}\gg\tilde{\rho}_{s}$).

With the above mentioned properties of $g$ we
choose as an example (the main results are independent of the particular $g$)
\begin{equation}
\label{eq:gfunction}
g({\bf r}_i(t))=\frac{1+\tanh\left[2\left(a-b\tilde{\rho}_{s}({\bf r}_i)+c\tilde{\rho}_{l}({\bf r}_i)\right)-1\right]}{2},
\end{equation}
where parameters $b$ and $c$ weight the importance of the short
and the long-range densities, respectively, and parameter $a$
gives the diffusivity of an individual when short and
long-range densities are equal and have the same weight. Notice
once again that $g \to 0$ if
$\tilde{\rho}_{s}\gg\tilde{\rho}_{l}$ and $g \to 1$ if
$\tilde{\rho}_{l}\gg\tilde{\rho}_{s}$.

\subsection*{Continuum description}

The  particle dynamics given by Eq. (\ref{langevin}) allows an
intensive numerical exploration. To complement the study and
obtain analytical results it is essential to have a simplified
continuum equation of the model, where the population is
described in terms of a collective variable: the local density
of individuals $\rho({\bf r})$. This equation can be derived
following Dean's approach \cite{dean1996langevin} from the
stochastic particle dynamics presented in the previous section,
which uses It\^{o} calculus. Considering a mean-field
approximation (i.e., neglecting fluctuations in the density) we
obtain for the mean particle density
\begin{equation}\label{density}
 \frac{\partial \rho({\bf r},t)}{\partial t}=
D_{0}\nabla^{2}\left[g(\tilde{\rho}_{s},\tilde{\rho}_{l})\rho({\bf r},t)\right],
\end{equation}
where
the mean long- and short-range densities are computed as
\begin{equation}
\tilde{\rho}_{\mu}({\bf r},t)=\int G_{\mu}({\bf r}-{\bf r'})\rho({\bf r'},t)d{\bf r'},
\end{equation}
where $G_{\mu}$, with $\mu\equiv s$ or $\mu\equiv l$, are the
short and long range kernel functions that define both
interaction scales. The kernel functions are normalized and
have units of inverse of area
\begin{eqnarray}\label{kernels}
 G_{\mu}(|{\bf r}|)=\left\{ \begin{array}{lcc}
                            \frac{1}{\pi R^{2}_{\mu}} & {\rm if} & |{\bf r}| \leq R_{\mu} \\
                            0 & {\rm if} & |{\bf r}| > R_{\mu}, \\
                           \end{array}
  \right.
\end{eqnarray}
$R_{\mu} \ (\mu\equiv s,r)$ define, as in the individual based approach,
the short and long interaction ranges, respectively.

\section*{Results}
\begin{figure}
\centering
\includegraphics[width=0.65\textwidth]{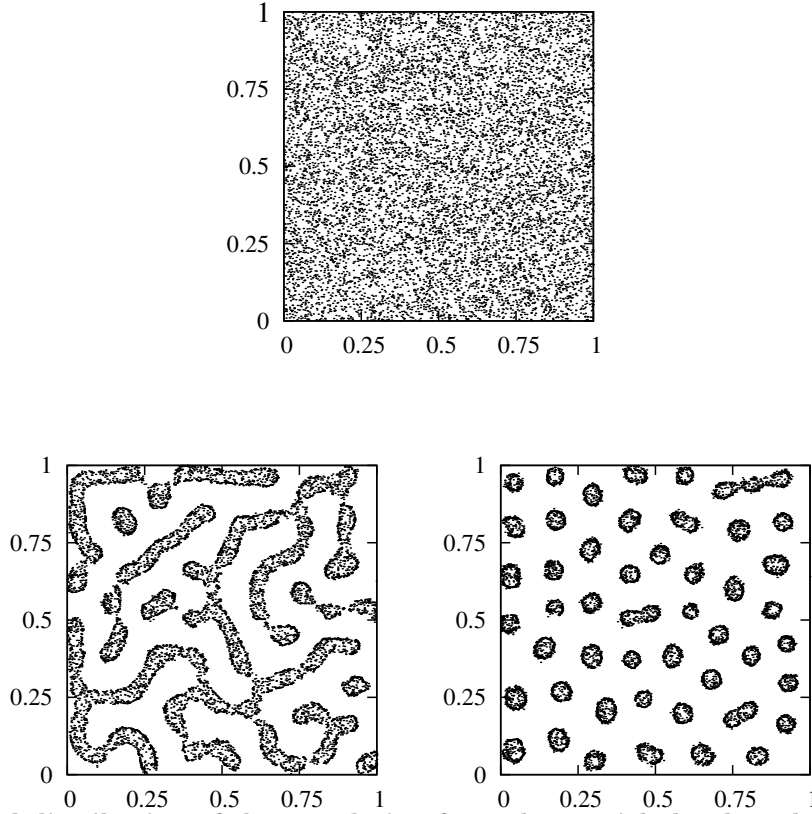}
\caption{{\bf Spatial distribution of the population from the particle-level model.}
Spatial distribution at long times of a population of $10^{4}$ individuals
using the dynamics of Eq.~(\ref{langevin}) with a short interaction range
$R_{s}=0.05$ and a long interaction length $R_{l}=0.1$. $D_{0}=10^{-4}$, $a=1$ in all the panels.
Every individual is plotted as a small black dot.
The system is a square area of lateral size $L=1$ with periodic boundary conditions.
Upper panel: $b=3.5\times10^{-4}$, $c=7.0\times10^{-4}$ (homogeneous distribution).
Left bottom panel: $b=8.5\times10^{-4}$, $c=7\times10^{-4}$ (labyrinth pattern). Right bottom
panel: $b=4.3\times10^{-4}$, $c=3.9\times10^{-4}$ (spot pattern).
Note the rings with higher density in the border.}
\label{fig2}
\end{figure}

A direct exploration of pattern formation in the model starts
from Monte Carlo numerical simulations of the individual-based
dynamics given by Eq.~\ref{langevin}. To unveil the
relationships between the two spatial scales that promote the
formation of spatial structures, we isolate in our analysis the
relative importance of the short and long-range densities
fixing all the parameters of the model ($R_s$, $R_l$, $D_0$,
$N$, $a$; see caption of Fig.~\ref{fig2} for details),
except $b$ and $c$, that weight the influence of $\tilde\rho_s$
and $\tilde\rho_l$ on the diffusivity.

Depending on the relationships between this pair of parameters
the population may show a homogeneous distribution (Fig.~
\ref{fig2}, top panel) or arrange developing spatial
aggregations (bottom  panels of Fig.~\ref{fig2}). Two
classes of patterns are observed: labyrinth distributions and
isolated spots \cite{Liu2013,VandeKoppel2008} arranged in a
hexagonal matrix. A relevant and singular feature is the shape
of the aggregations, with most of the individuals clumped in
the borders of the cluster and an almost empty inner area.
Similar ring-like structures have been previously reported in
plant ecology and studied with models based on mechanisms very
different form ours, but that share with our approach the
presence of competitive and facilitative interactions
\cite{fernandez2014strong,Sheffer2011mechanisms}.

\begin{figure}
\centering
\includegraphics[width=0.65\textwidth]{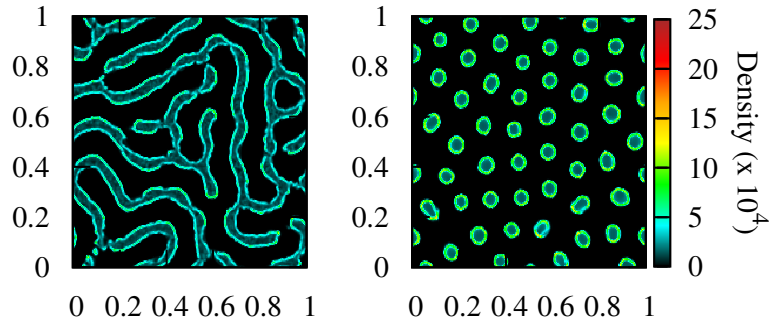}
\caption{{\bf Solutions of the continuous density equation.}
Long time solution of Eq.~(\ref{density}) with a short interaction range
$R_{s}=0.05$ and a long interaction length $R_{l}=0.1$. $D_{0}=10^{-4}$, $a=1$ and
density $\rho_{0}=10^{4}$ in all the panels.
An Euler algorithm was implemented and integration performed over a square area
with lateral size $L=1$ and periodic boundary conditions.
Left panel: $b=8.5\times10^{-4}$, $c=7\times10^{-4}$ (labyrinth pattern).
Right panel: $b=4.3\times10^{-4}$, $c=3.9\times10^{-4}$ (spot pattern).}
\label{fig3}
\end{figure}

\begin{figure}
\centering
\includegraphics[width=0.65\textwidth]{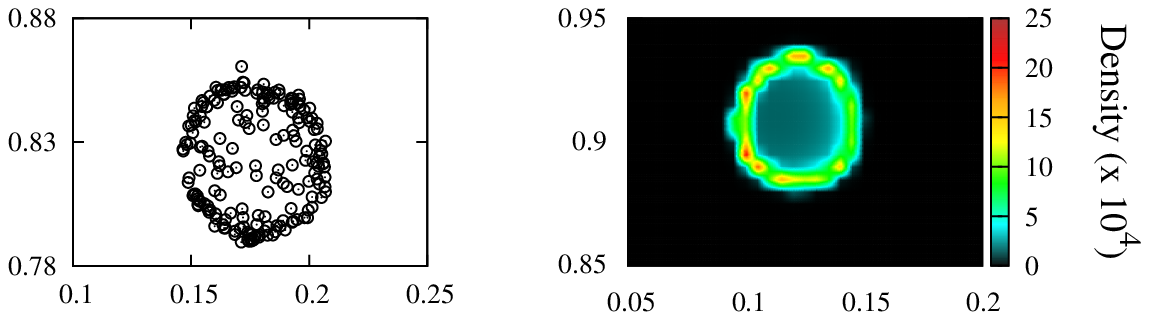}
\caption{{\bf Comparison of a single ring-like structure in both approaches.}
Detailed distribution of the individuals within one of the groups of the spotted
pattern in the discrete model (Left) and of the density in one of the patches in the solution
of the continuous equation (Right).
Parameters: $b=4.3\times10^{-4}$, $c=3.9\times10^{-4}$, $D_{0}=10^{-4}$, $a=1$, $R_{l}=0.1$ and $R_{s}=0.05$ in all the panels.}
\label{fig4}
\end{figure}

A deeper understanding of the pattern formation dynamics can be
addressed using the continuum description given by
Eq.~(\ref{density}). To corroborate the correspondence between
the individual based description by  Eq.~(\ref{langevin}) and
the continuous approach in terms of Eq.(~\ref{density}), we
numerically integrate Eq.~(\ref{density}). Kernels are fixed as
given by Eq.~(\ref{kernels}) and the parameters take the same
values as in Fig.~\ref{fig2} to allow a direct
comparison with the discrete simulations (see caption of Fig.~
\ref{fig3} for details). The laberynth and spot patterns showed in
Fig.~\ref{fig3} exhibit a good agreement with the
distributions of Fig.~\ref{fig2} resulting from the
stochastic particle dynamics. In particular, details of hollow
clusters for both micro and macro descriptions are plotted in
Fig.~\ref{fig4}. The distribution of the particles within
the clusters is a particularly interesting question that will
be discussed later in this section.

\begin{figure}
\centering
\includegraphics[width=0.65\textwidth]{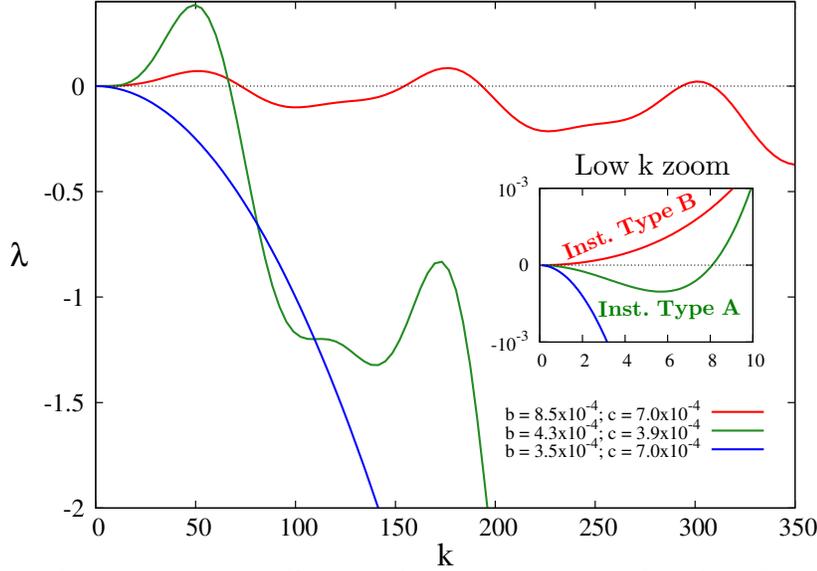}
\caption{{\bf Perturbation growth rate.} Perturbation growth rate as a function of the wavenumber, Eq.~(\ref{dispersion}),
for different values of the parameters $b$ and $c$. $R_{s}=0.05$, $R_{l}=0.1$, $D_{0}=10^{-4}$ and $a=1$.}
\label{fig5}
\end{figure}

We continue with the analytical approach performing a linear
stability analysis of Eq.~(\ref{density}). We note that the
homogeneous distribution of the $N$ individuals in the box of
size $L$, i.e. $\rho({\bf r},t)=\rho_0=N/L^2$ always provides a
stationary solution to such equation. The stability of this
homogeneous distribution is checked by adding a small
perturbation to it, so that $\rho({\bf
r},t)=\rho_{0}+\epsilon\psi({\bf r},t)$ ($\epsilon\ll 1$).
Inserting this into Eq.~(\ref{density}) we find that the
perturbation growth rate of $\psi\propto \exp({\bf k}\cdot {\bf
r} + \lambda t$) is given by
\begin{equation}\label{dispersionlambda}
\lambda({\bf k})=-\frac{D_{0}}{2}\left(1+\tanh\gamma+\frac{2c \rho_0 \hat{G}_{l}({\bf k})-2b \rho_0 \hat{G}_{s}({\bf k})}{\cosh^{2}\gamma}\right){\bf k}^{2},
\end{equation}
where $\gamma=2(a-b \rho_0 +c\rho_0)-1$. $\hat{G}_{s}({\bf k})$
and $\hat{G}_{l}({\bf k})$ are the Fourier transforms of the
short-range and the long-range kernels, respectively. Given the
choice made for the kernels (Eq. (\ref{kernels})), the Fourier
transforms are
\begin{equation} \label{dispersion}
\hat{G}_{\mu}({\bf k})=2\frac{J_{1}({\bf k}R_{\mu})}{{|\bf k|}R_{\mu}}, \
\end{equation}
where $\mu=s$ or $\mu=l$, and $J_{1}$ is the first order Bessel
function. The homogeneous distribution is unstable and then
patterns would appear if the maximum of the growth rate (i.e.,
of the most unstable mode), $\lambda({\bf k}_{c})$, is
positive, which means that the perturbation of periodicity
$2\pi/|{\bf k}_c|$ grows with time. $\lambda$ is showed for
different values of the parameters $b$ and $c$ in
Fig.~\ref{fig5}. Depending on the value of $b$ and
$c$ the model shows two different types of instabilities.
Instability A has stable low wavenumbers (green curve in
Fig.~\ref{fig5}, see inset) that prevent the clusters
to grow. The characteristic wavelength of the pattern is well
defined around $k_c = 49.52$. On the other hand an instability
of type B has a band of unstable modes starting at $k=0$, which
could allow the clusters to experience some coarsening in time.
We observe that labyrinthic structures are formed by this type
B instability.

\begin{figure}
\centering
\includegraphics[width=0.45\textwidth]{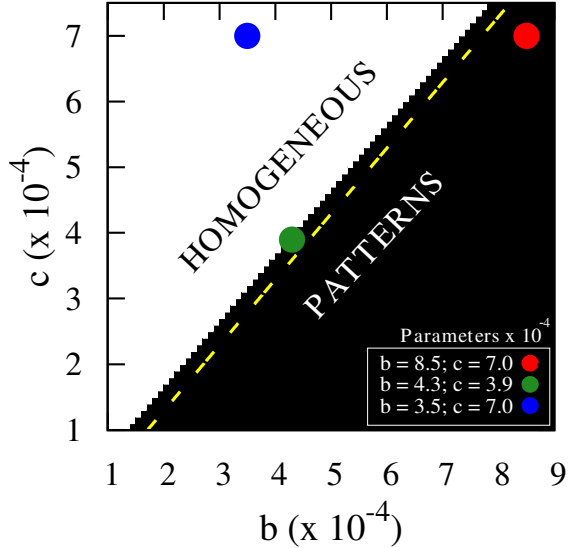}
\caption{{\bf Phase diagram of the continuous approach.} Parameter space of the continuous model where the regions of patterns and homogeneous solutions have been 
identified using the perturbation growth. $R_{s}=0.05$, $R_{l}=0.1$, $D_{0}=10^{-4}$ and $a=1$.
The yellow dashed line shows the transition from Instability A (above the line) to Instability B (below the line).}
\label{fig6}
\end{figure}

Evaluating the perturbation growth rate in Eq.
(\ref{dispersionlambda}) with Eq. (\ref{dispersion}), we may
compute the phase diagram of the model (see
Fig.~\ref{fig6}) for parameters $b$ and $c$ that gives
information about the final spatial distribution of the system,
homogeneous or patterned. The reduction of the diffusivity at
high short-range densities is the responsible of the formation
of patterns since clusters appear when $b>c$, that is when
$\rho_s$ is more relevant for the dynamics than $\rho_l$. On
the other hand, considering the effect of the long-range
density alone on the diffusivity, the system shows homogenous
distributions regardless of the value of $c$ when $b=0$. These
are expected results since high values of the short-range
density reduce the mobility of the individuals promoting
clustering, while high values of the long-range density enhance
longer displacements in the population, thus leading to
homogenous distributions. However, the instability caused by a
diffusivity reduction is enhanced by the presence of the
$\rho_l$ dependence because animals in-between two aggregations
make longer displacements that allow them to reach a group. A
similar argument has been used to explain the formation of
clusters of species \cite{pigolotti2007species} and vegetation
\cite{martinez2013vegetation} in systems that only present
long-range competitive interactions.

In addition, the boundary between both types of instabilities
(A for hexagonal clusters, and B for labyrinthic patterns) is
given by a change in the sign of the second derivative of the
perturbation growth rate at $k=0$. It is represented in
Fig.~\ref{fig6} by the yellow dashed line resulting from
numerically evaluating
\begin{equation}
 \lambda''(k)\rvert_{k=0}=\frac{-D_0}{2}\left(1+\frac{2(c-b)}{\cosh^{2}\gamma}+\tanh\gamma\right)=0.
\end{equation}

\begin{figure}
\centering
\includegraphics[width=0.65\textwidth]{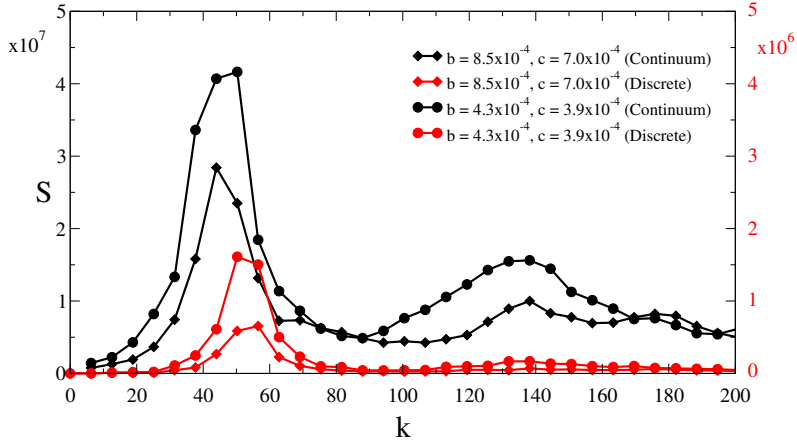}
\caption{{\bf Structure functions.} Structure function of the patterns 
obtained with the continuous and the discrete model for the case of labyrinthic and spotted patterns.}
\label{fig7}
\end{figure}

The typical scale of the pattern, that is, the distance between
aggregates, can be studied with the structure function (Fig.~\ref{fig7}). It can
be computed for both the patterns of particles and the density
distribution. In the first case it is $S_{d}({\bf
k})=\left\langle \left|\frac{1}{N}\sum_{j=1}^{N}{\rm e}^{i{\bf
k}\cdot{\bf r}_{j}}\right|^2 \right\rangle$, where ${\bf
r}_{j}$ is the position vector of particle $j$, ${\bf k}$ is a
two-dimensional wave vector with modulus $k$, and the average
indicates a spherical average over the wave vectors with
modulus $k$ and in time. In the continuous approach, the
structure function is calculated as the modulus of the spatial
Fourier transform of density field, averaged spherically and in
time. Note that both quantities are related but not identical,
and their first maximum, $k_c$,  allows to compute the typical
distance between clusters  $d = 2\pi/k_c$. For the spotted
patterns $k_{c}=50.24$ (discrete) and $k_c=49.52$ (continuum)
so that $d \approx 0.125-0.126$. Regarding the case of the
labyrinth pattern (central panel of Figs. \ref{fig2} and
\ref{fig3}), $k_{c}=56.52$ (discrete) and $k_c =51.31$
(continuum), so that the typical distance between aggregates is
$d \approx 0.11$.

\begin{figure}
\centering
\includegraphics[width=0.65\textwidth]{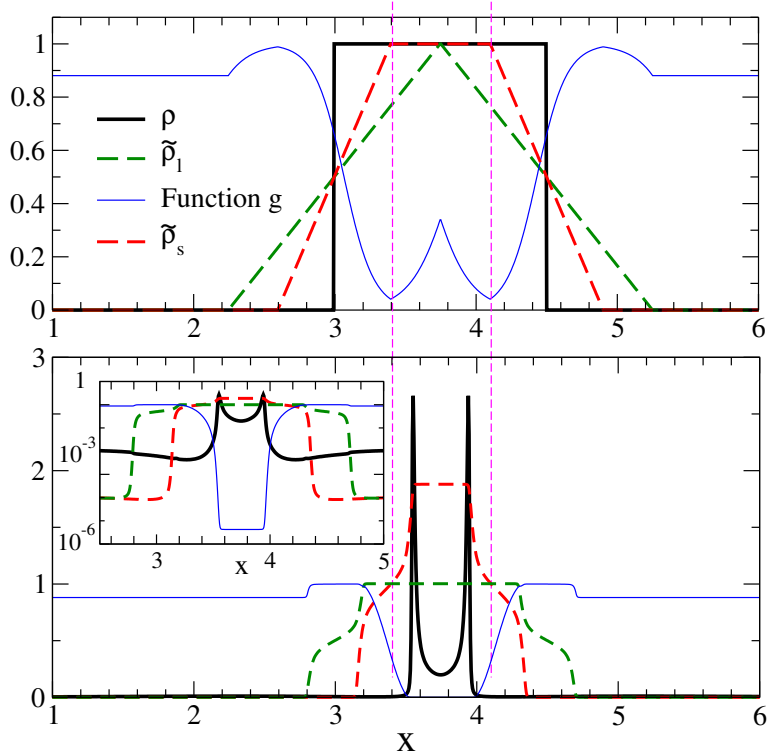}
\caption{{\bf Formation of ring-like structures in the 1D model}. Evolution of the 1D version of the model
starting from an initial condition for $\rho(x,t=0)$ consisting on a pulse
of height unity and length $2R_l$ (displayed in the upper panel). The
legend indicates the quantity represented by
the different lines. The two vertical lines indicate the minima of the function $g(x)$ (i.e.
the diffusion coefficient in units of $D_0$) at $t=0$ where particles
will initially tend to accumulate. The bottom panel represent the same
quantities (although $\rho(x,t)$ has been divided by a factor 5 to fit into the same scale as the other
curves) after a very long integration time ($t=6\times 10^5$). A double-peak structure
has developed. The inset displays this long-time
configuration in logarithmic scale, showing that $g(x)\approx 4\times 10^{-6}$ in the central region.
Parameters: $a=1$, $b=3.33$, $c=2.67$, $D_0=10^{-4}$, $R_l=0.75$, $R_s=0.4$,
$N=\int dx \rho(x,t=0)=1.5$}.
\label{fig8}
\end{figure}

\begin{figure}
\centering
\includegraphics[width=0.45\textwidth]{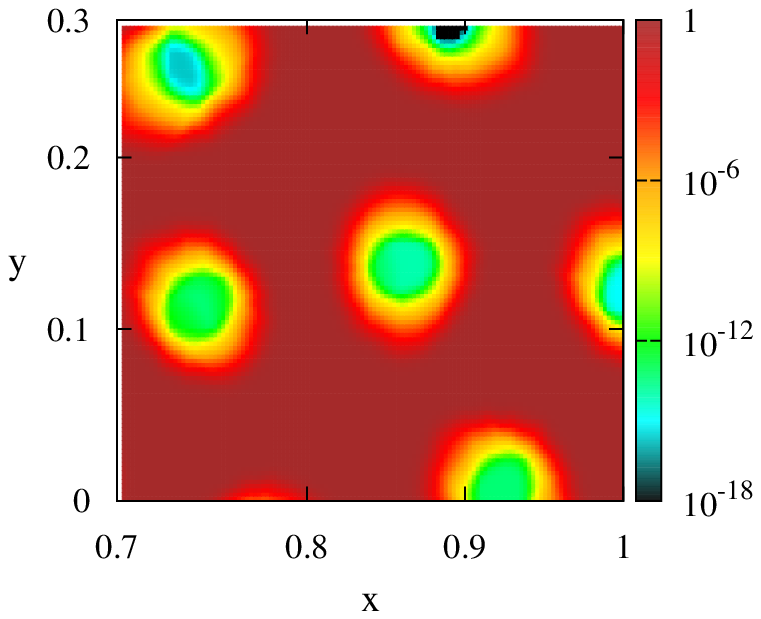}
\caption{{\bf Diffusion field in the 2D model.} Numerical computation of the function $g$ as defined in Eq.~(\ref{eq:gfunction})
from the spot pattern showed in Fig.~\ref{fig3}. $g$ is extremely small inside the clusters, but not zero.
Parameters: $R_{s}=0.05$, $R_{l}=0.1$, $D_{0}=10^{-4}$, $a=1$, $\rho_{0}=10^{4}$,
$b=4.3\times10^{-4}$, $c=3.9\times10^{-4}$. Zoom over the right lower corner of the pattern.}
\label{fig9}
\end{figure}

As it was stated before, the ring-like shape of the clusters
deserves further consideration. To go deeper into this question
we use the one-dimensional version of the model starting from
an initial condition consisting of a single pulse of height
unity (top panel of Fig.~\ref{fig8}). The mean
nonlocal densities $\tilde{\rho}_{s}$ (dashed red line) and
$\tilde{\rho}_{l}$ (dashed green line) can be easily obtained
and lead to a diffusivity which in units of $D_0$ is the
function $g$, with two minima where particles will tend
initially to clump (magenta vertical dashed lines in Fig.~
\ref{fig8}). As time advances a two-peak distribution
establishes, which is the one-dimensional analogue of the
two-dimensional rings observed before. This double peak, of a
spatial size close to $R_s$, persists for extremely long times.
However the inset in the bottom panel of
Fig~\ref{fig8} shows that the diffusion coefficient
in between the two peaks takes a nearly constant value which is
very small but not zero ($g(x)=D(x)/D_0 \approx 4\times
10^{-6}$). This implies that at still longer times (of the
order of $R_s^2/D\approx 4\times 10^5$ after the time displayed
in the bottom panel of Fig~\ref{fig8}) particles will
diffuse between the two peaks, replacing them by a homogeneous
distribution. The same will occur in two dimensions, since as
showed in Fig.~\ref{fig9}, the diffusion
coefficient in the two-dimensional system is also homogeneous
(but very small) inside the clusters so that at extremely long
times the pattern of hollow clusters of Fig.~\ref{fig3}
will be replaced by homogeneous clusters. Thus the ring
structures seem to be a very-long lived transient state. They
will disappear faster if the prescription in
Eq.~(\ref{eq:gfunction}) for $g$ is changed by another
functional form with higher minimum values. Alternatively, for
a choice such that $g(x)$ is strictly zero for $\tilde{\rho_s}
\gg \tilde{\rho_l}$ then the rings will persists for infinite
time as stationary structures.

\section*{Discussion}
We have studied how the combination of a short-range inhibition
and a long-range activation in individual dispersal may
influence the long-time spatial distribution of a population,
which ranges from homogeneous to labyrinth and spot patterns
depending on the relative weights of each mechanism. This type
of behavior has been observed in mussel beds
\cite{VandeKoppel2005,VandeKoppel2008,DeJager2011} where
individuals tend to clump at short distance as a defensive
strategy while competition for resources acts at a larger
scale.

Pattern formation arises as a consequence of the interplay
between inhibition and activation acting at different spatial
scales that makes the spatially homogeneous state to lose its
stability. Resulting patterns show not only an inhomogeneous
distribution of the population at a system level but also a
non-uniform distribution of the individuals within each cluster.
For the time scales discussed here ring-like structures are
formed, with most of the particles at the borders of the
groups. This point has been studied from a simplified 1D
situation starting from an initial density given by a step
function. Beyond the limits of the profile the
nonlocal long-range densty is higher than the nonlocal short-range density.
However, due to their different slopes, this situation is reversed
and the short-range density becomes higher than the long-range one.
This leads to the formation of annular structures.
This mechanism will act for any kind of initial condition
wherever there is a region where eventually the density is
higher than in the rest of the system. Whether the rings will
homogenize at very long times or rather they will remain stable
depends on the details of the small-diffusion part of the
density-dependent diffusivity.

The particular shape of the structures depends on the relative
importance of the short and long-range mean densities, weighted
by the values of the parameters $b$ and $c$. The first is the
responsible of the formation of aggregates, so the model gives
homogeneous distributions when this scale tends to zero
($R_{s}\rightarrow 0$ or equivalently $b=0$). The larger one
enhances the formation of groups. Individuals that do not
belong to any group are surrounded by low densities in their
close neighborhoods, but still can be in very far-populated
regions. In these cases their movement has a larger
diffusivity, so longer displacements are possible, increasing
the probability of finding a group in a shorter time. A
combination of both, a short- and a long-range dependence
mobility, is an optimal mechanism to promote the formation of
groups. In addition, the long-range competition  stabilizes the
ring-like structures since it avoids the formation of highly
packed groups in a small area.

The generality of the model, a nonlinear diffusion equation
with two nonlocal interaction scales that enhance and inhibit
animal mobility, allows its application to a wide variety of
ecological situations with these two ingredients. Moreover, our
mathematical scheme shows a sequence of patterns that has been
previously reported in natural systems such as mussel beds
\cite{DeJager2011}
(isolated spots spatially
arranged at random can also be observed for
slightly different setups
of the model, for example, changing the 
hyperbolic tangent function in Eq. (\ref{eq:gfunction})).
In the case of mussels, long-range activation of dispersal arises from resources competition. 
Individuals would tend to escape from regions already colonized by other groups. Nevertheless 
they will remain inside a small group if at this 
smaller scales clustering provides some advantage such as protection against wave stress. 
Protection against predators is also a widespread benefit of clustering in groups. 
Within our approach, we recover spatial structures both in a
stochastic and a deterministic description of the problem,
suggesting that they are a result of the interplay between the
two types of interactions with fluctuations playing a secondary
role. Remarkably, our results bear similarities with results on
vegetation patterns and fairy circles in arid regions
\cite{fernandez2014strong,Sheffer2011mechanisms,juergens2013biological}
which arise from very different mechanisms, but have in common
with our case the presence of competitive and facilitative
interactions. We hope that our studies help the development of
new mathematical models and more precise understanding of those
situations where spatial distributions similar to the ones
presented here are observed.

\section*{Acknowledgments}
We acknowledge fruitful discussions with Prof. Daniel Walgraef.
RMG was suportted by the JAEPredoc program of Consejo Superior de
Investigacion Cientificas (CSIC), CM was supported by the IFISC-Intro program of Instituto de
Fisica Interdisciplinar y Sistemas Complejos. EHG and CL are supported by
Ministerio de Economia y Competitividad and Fondo Europeo de Desarrollo Regional
through grants No. FIS2012-30634 (INTENSE@COSYP) and No. CTM2012-39025-C02-01
(ESCOLA).

\nolinenumbers

\end{document}